\documentclass[prc,preprint,showpacs,showkeys,nofootinbib]{revtex4}%
\usepackage{graphicx}
\usepackage{bm}
\usepackage{epsfig}
\usepackage{amsmath}
\usepackage{amsfonts}
\usepackage{amssymb}%

\begin{document}
\title{Excitation of soft dipole modes in electron scattering}
\author{C.A. Bertulani$^{1,2}$}
\email{bertulanica@ornl.gov} \affiliation{$^{1}$Department of
Physics and Astronomy University of Tennessee Knoxville, Tennessee
37996, USA \\
$^{2}$Physics Division Oak Ridge National Laboratory P.O. Box 2008
Oak Ridge, Tennessee 37831, USAA}

\begin{abstract}
The excitation of soft dipole modes in light nuclei via inelastic
electron scattering is investigated. I show that, under the proposed
conditions of the forthcoming electron-ion colliders, the scattering
cross sections have a direct relation to the scattering by real
photons. The advantages of electron scattering over other
electromagnetic probes is explored. The response functions for
direct breakup are studied with few-body models. The dependence upon
final state interactions is discussed. A comparison between direct
breakup and collective models is performed. The results of this
investigation are important for the planned electron-ion colliders
at the GSI and RIKEN facilities.

\end{abstract}
\maketitle


\section{Introduction}

Reactions with radioactive beams have attracted great experimental and
theoretical interest during the last two decades \cite{BHM02}. Progresses of
this scientific adventure were reported on measurements of nuclear sizes
\cite{Ta85}, the use of secondary radioactive beams to obtain information on
reactions of astrophysical interest \cite{BBR86,BB88}, fusion reactions with
neutron-rich nuclei \cite{TS91,Hus92}, tests of fundamental interactions
\cite{Har97}, dependence of the equation of state of nuclear matter upon the
asymmetry energy \cite{DLL02}, and many other research directions. Studies of
the structure and stability of nuclei with extreme isospin values provide new
insights into every aspect of the nuclear many-body problem. In neutron-rich
nuclei far from the valley of $\beta$-stability, in particular, new shell
structures occur as a result of the modification of the effective nuclear
potential. Neutron density distributions become very diffuse and the
phenomenon of the evolution of the neutron skin and, in some cases, the
neutron halo have been observed.

New research areas with nuclei far from the stability line will
become possible with newly proposed experimental facilities. Among
these we quote the FAIR facility at the GSI laboratory in Germany.
One of the projects for this new facility is the study of electron
scattering off unstable nuclei in an electron-ion collider mode
\cite{Haik05}. A similar proposal exists for the RIKEN\ laboratory
facility in Japan \cite{Sud01}. By means of elastic electron
scattering, these facilities will become the main tool to probe the
charge distribution of unstable nuclei \cite{Ant05,Ber06}. This will
complement studies of matter distribution which have been performed
in other radioactive beam facilities using hadronic probes.
Inelastic electron scattering will test the nuclear response to
electromagnetic fields.

These facilities will provide accurate measurements of many nuclear
properties of unstable nuclei. The reason is that electron
scattering is a very clean probe. Its electromagnetic interaction
with the nucleus is well understood. Inelastic electron scattering
can also be very well described in the Born approximation. Higher
order processes are only relevant for the distortion of the electron
wavefunctions, affecting mostly electron scattering on heavy nuclei.

Up to now, the electromagnetic response of unstable nuclei far from the
stability line has been studied with Coulomb excitation of radioactive beams
impinging on a heavy target \cite{BB88}. This method has been very useful in
determining the electromagnetic response in light nuclei \cite{BCH93}. For
neutron-rich isotopes \cite{Lei01} the resulting photo-neutron cross sections
are characterized by a pronounced concentration of low-lying $E1$ strength.
The onset of low-lying $E1$ strength has been observed not only in exotic
nuclei with a large neutron excess, but also in stable nuclei with moderate
proton-neutron asymmetry. The problem with such experiments is that the probe
is not very clean. It is well known that the nuclear interaction between
projectile and target as well as the long range Coulomb distortion of the
energy of the fragments interacting with the target (see, e.g. ref.
\cite{Ber05}) are problems of a difficult nature. The nuclear response probed
with electron does not suffer from these inconveniences.

The interpretation of the low-lying $E1$ strength in neutron-rich
nuclei engendered a debate: are these \textquotedblleft soft dipole
modes\textquotedblright\ just a manifestation of the loosely-bound
character of light neutron-rich nuclei, or are they a manifestation
of the excitation of a resonance? \cite{Iek93,Sac93,Sag95,Hus96}. As
far as I know, there has not been a definite answer to this simple
question. This apparently innocuous question has nonetheless become
the center of a even more widespread debate. It is believed that the
weak binding of outermost neutrons gives rise to a direct break up
of the nucleus and a consequent concentration of the electromagnetic
response at low energies. The same weak binding can also lead to
soft collective modes. In particular, the pygmy dipole resonance
(PR), i.e. the resonant oscillation of the weakly-bound neutron
mantle against the isospin saturated proton-neutron core. Its
structure, however, remains very much under discussion. The
electromagnetic response of light nuclei, leading to their
dissociation, has a direct connection with the nuclear physics
needed in several astrophysical sites \cite{BBR86,BB88,Ber05}. In
fact, it has been shown \cite{Go98} that the existence of pygmy
resonances have important implications on theoretical predictions of
radiative neutron capture rates in the r-process nucleosynthesis,
and consequently to the calculated elemental abundance distribution
in the universe.

In this work I study the general features of inelastic electron
scattering off light nuclei, in particular their response in the
continuum. An assessment of the theory of inelastic electron
scattering appropriate for the conditions of electron-ion colliders
is presented in section 2. Special emphasis is put on the connection
of electron scattering and the scattering by real photons, which
will be useful to relate electron scattering and Coulomb
dissociation measurements. Section 3 deals with the nuclear response
within two and three-body models and their dependence upon final
state interactions. Section 4 discusses the aspects of low energy
collective modes in halo nuclei and their connection with the
response obtained with few-body models. The summary and conclusions
will be presented in section 5.

\section{Inelastic Electron Scattering}

In the plane wave Born approximation (PWBA) the cross section for inelastic
electron scattering is given by \cite{Ba62,EG88}%
\begin{align}
\frac{d\sigma}{d\Omega}  &  =\frac{8\pi e^{2}}{\left(  \hbar c\right)  ^{4}%
}\left(  \frac{p^{\prime}}{p}\right)  \sum_{L}\left\{  \frac{EE^{\prime}%
+c^{2}\mathbf{p\cdot p}^{\prime}+m^{2}c^{4}}{q^{4}}\left\vert F_{ij}\left(
q;CL\right)  \right\vert ^{2}\right. \nonumber\\
&  \left.  +\frac{EE^{\prime}-c^{2}\left(  \mathbf{p\cdot q}\right)  \left(
\mathbf{p}^{\prime}\cdot\mathbf{q}\right)  -m^{2}c^{4}}{c^{2}\left(
q^{2}-q_{0}^{2}\right)  ^{2}}\left[  \left\vert F_{ij}\left(  q;ML\right)
\right\vert ^{2}+\left\vert F_{ij}\left(  q;EL\right)  \right\vert
^{2}\right]  \right\}  \label{PWBA}%
\end{align}
where $J_{i}$ $\left(  J_{f}\right)  $ $\ $is the initial (final)
angular momentum of the nucleus, $\left(  E,\mathbf{p}\right)  $ and
($E^\prime ,\mathbf{p}^{\prime}$) are the initial and final energy
and momentum of the electron, and $\left(  q_{0},\mathbf{q}\right)
=\left(\frac{  (E-E^{\prime})}{\hbar c},\frac{\left(
\mathbf{p-p}^{\prime}\right)} {\hbar}\right)  $ is the energy and
momentum transfer in the reaction. $F_{ij}\left(  q;\Pi L\right)  $
are form factors for momentum transfer $q$ and for Coulomb ($C$),
electric ($E$) and magnetic ($M$) multipolarities, $\Pi=C,E,M$,
respectively.

Here we will only treat electric multipole transitions. Moreover, we will
treat low energy excitations such that $E,E^{\prime}\gg\hbar cq_{0}$, which is
a good approximation for electron energies \ $E\simeq500$ MeV and small
excitation energies $\Delta E=\hbar cq_{0}\simeq1-10$ MeV. These are typical
values involved in the dissociation of nuclei far from the stability line.

Using the Siegert's theorem \cite{Sie37,Sa51}, one can show that
the Coulomb and electric form factors in eq.
\ref{PWBA} are proportional to each other. Moreover, for very forward
scattering angles ($\theta\ll1$) the PWBA cross section, eq.
\ref{PWBA}, can be rewritten as
\begin{equation}
\frac{d\sigma}{d\Omega dE_{\gamma}}=\sum_{L}\frac{dN_{e}^{(EL)}\left(
E,E_{\gamma},\theta\right)  }{d\Omega dE_{\gamma}}\ \sigma_{\gamma}%
^{(EL)}\left(  E_{\gamma}\right)  , \label{EPA}%
\end{equation}
where $\sigma_{\gamma}^{(EL)}\left(  E_{\gamma}\right)  $, with $E_{\gamma
}=\hbar cq_{0}$,\ is the photo-nuclear cross section for the $EL$%
-multipolarity, given by \cite{BB88}%
\begin{equation}
\sigma_{\gamma}^{(EL)}\left(  E_{\gamma}\right)  =\frac{\left(  2\pi\right)
^{3}\left(  L+1\right)  }{L\left[  \left(  2L+1\right)  !!\right]  ^{2}%
}\left(  \frac{E_{\gamma}}{\hbar c}\right)  ^{2L-1}\frac{dB\left(  EL\right)
}{dE_{\gamma}}. \label{sigphoto}%
\end{equation}
In the long-wavelength approximation, the response function, $dB\left(
EL\right)  /dE_{\gamma},$ in eq. \ref{sigphoto} is given by
\begin{equation}
\frac{dB\left(  EL\right)  }{dE_{\gamma}}=\frac{\left\vert \left\langle
J_{f}\left\Vert Y_{L}\left(  \widehat{\mathbf{r}}\right)  \right\Vert
J_{i}\right\rangle \right\vert ^{2}}{2J_{i}+1}\left[  \int_{0}^{\infty
}dr\ r^{2+L}\ \ \delta\rho_{if}^{\left(  EL\right)  }\left(  r\right)
\right]  ^{2}w\left(  E_{\gamma}\right)  , \label{photo}%
\end{equation}
\newline where $w\left(  E_{\gamma}\right)  $ is the density of final
states (for nuclear excitations into the continuum) with energy $E_{\gamma
}=E_{f}-E_{i}$. The
the transition density $\delta\rho_{if}^{\left(  EL\right)  }\left(  r\right)
$ will depend upon the nuclear model adopted.

For $L\ge 1$ one obtains from eq. (1) that
\begin{align}
\frac{dN_{e}^{(EL)}\left(  E,E_{\gamma},\theta\right)  }{d\Omega dE_{\gamma}}
&  =\frac{4L}{L+1}\frac{\alpha}{E}\left[  \frac{2E}{E_{\gamma}}\sin\left(
\frac{\theta}{2}\right)  \right]  ^{2L-1}\nonumber\\
&  \times\frac{\cos^{2}\left(  \theta/2\right)  \sin^{-3}\left(
\theta/2\right)  }{1+\left(  2E/M_{A}c^{2}\right)  \sin^{2}\left(
\theta/2\right)  }\left[  \frac{1}{2}+\left(  \frac{2E}{E_{\gamma}}\right)
^{2}\frac{L}{L+1}\sin^{2}\left(  \frac{\theta}{2}\right)  +\tan^{2}\left(
\frac{\theta}{2}\right)  \right]  . \label{EPAE}%
\end{align}

One can also define a differential cross section integrated over angles. Since
$\sigma_{\gamma}^{(EL)}$ does not depend on the scattering
angle, this can be obtained from eq. \ref{EPAE} by integrating $dN_{e}^{(EL)}
/d\Omega dE_{\gamma}$ over angles, from $\theta_{\min}=E_{\gamma}/E$ to
a maximum value $\theta_{m}$, which depends upon
the experimental setup.

Eqs. \ref{EPA}-\ref{EPAE} show that under the conditions of the
proposed electron-ion colliders, electron scattering will offer the
same information as excitations induced by real photons. The
reaction dynamics information is contained in the virtual photon
spectrum of eq. \ref{EPAE}, while the nuclear response dynamics
information will be contained in eq. \ref{photo}. This is akin to a
method developed long time ago by Fermi \cite{Fe24} and usually
known as the Weizsaecker-Williams method \cite{We34}. The quantities
$dN_{e}^{(EL)}/d\Omega dE_{\gamma}$ can be interpreted as the number
of equivalent (real) photons incident on the nucleus per unit
scattering angle $\Omega$ and per unit photon energy $E_{\gamma}.$
Note that $E0$ (monopole) transitions do not appear in this
formalism. As immediately inferred from eq. \ref{photo}, for $L=0$
the response function $dB\left( EL\right) /dE_{\gamma}$ vanishes
because the volume integral of the transition density also vanishes
in the long-wavelength approximation. But for larger scattering
angles the Coulomb multipole matrix elements ($CL$) in eq.
\ref{PWBA} are in general larger than the electric ($EL$)
multipoles, and monopole transitions become relevant \cite{Sch54}.

\begin{figure}[tb]
\begin{center}
\includegraphics[
height=2.8945in,
width=3.1375in
]{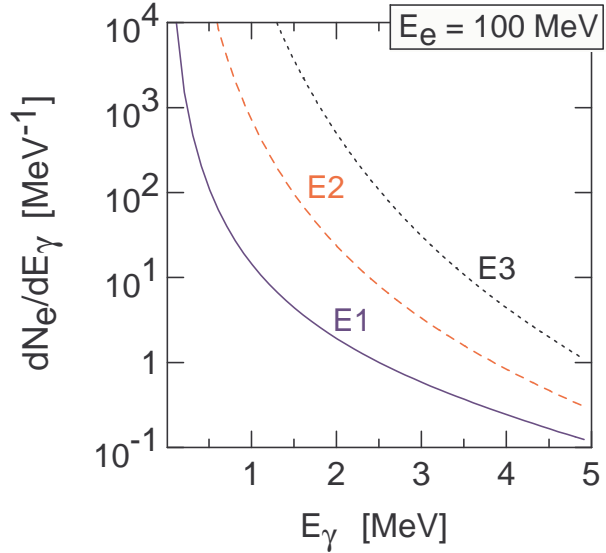}
\end{center}
\caption{(Color online) Virtual photon spectrum for the $E1$, $E2$
and $E3$ multipolarities in electron scattering off arbitrary nuclei
at $E_{e}=100$ MeV and maximum
scattering angle of 5 degrees.}%
\label{dnde1}%
\end{figure}

In figure \ref{dnde1} we show the virtual photon spectrum for the $E1$, $E2$
and $E3$ multipolarities for electron scattering off arbitrary nuclei at
$E_{e}=100$ MeV. These spectra have been obtained
assuming a maximum scattering angle of 5 degrees. An evident feature
deduced from this
figure is that the spectrum increases rapidly with decreasing energies. Also,
at excitation energies of 1 MeV, the spectrum yields the ratios $dN_{e}%
^{(E2)}/dN_{e}^{(E1)}\simeq500$ and $dN_{e}^{(E3)}/dN_{e}^{(E2)}\simeq100$.
However, although $dN_{e}^{(EL)}/dE_{\gamma}$ increases with the multipolarity
$L$, the nuclear response decreases rapidly with $L$, and $E1$ excitations
tend to dominate the reaction. For larger electron energies the ratios
$N^{(E2)}/N^{(E1)}$ and $N^{(E3)}/N^{(E1)}$ decrease rapidly.

\begin{figure}[ptb]
\begin{center}
\includegraphics[
height=2.911in,
width=3.2007in
]{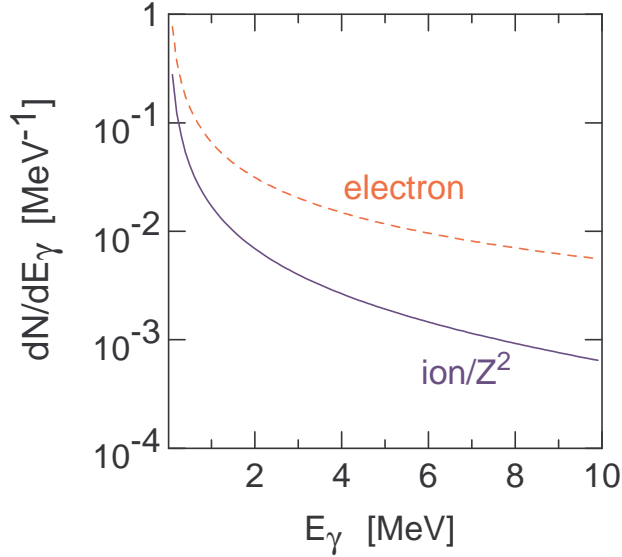}
\end{center}
\caption{(Color online) Comparison between the virtual photon
spectrum of 1 GeV electrons (dashed line), and the spectrum
generated by a 1 GeV/nucleon heavy ion projectile (solid line) for
the $E1$ multipolarity, as a function of the photon energy. The
virtual photon spectrum for the ion has been divided by the
square of its charge number.}%
\label{fig2}%
\end{figure}

Note that a similar relationship as eq. \ref{EPA} also exists for Coulomb
excitation \cite{BB88} in heavy ion scattering.
In figure \ref{fig2} we show a comparison between the $E1$ virtual photon
spectrum, $dN_{e}/dE_{\gamma}$, of 1 GeV electrons with the spectrum generated
by 1 GeV/nucleon heavy ion projectiles. In the case of Coulomb excitation, the
virtual photon spectrum was calculated in ref. \cite{BB88}, eq. 2.5.5a. For
simplicity, we use for the strong interaction distance $R=10$ fm. The spectrum
for the heavy ion case is much larger than that of the electron for large
projectile charges. For $^{208}$Pb projectiles it can be of the order of 1000
times larger than that of an electron of the same energy. As a natural
consequence, reaction rates for Coulomb excitation are larger than for
electron excitation. But electrons have the advantage of being a clean
electromagnetic probe, while Coulomb excitation at high energies needs a
detailed theoretical analysis of the data due to contamination by nuclear
excitation. As one observes in figure \ref{fig2}, the virtual spectrum for the
electron contains more hard photons, i.e. the spectrum decreases slower with
photon energy than the heavy ion photon spectrum. This is because, in both
situations, the rate at which the spectrum decreases depends on the ratio of
the projectile kinetic energy to its rest mass, $E/mc^{2},$ which is much
larger for the electron ($m=m_{e}$) than for the heavy ion ($m=$ nuclear mass).

To obtain an effective luminosity per unit energy, the equivalent
photon number is multiplied by the experimental luminosity,
$L_{eA}$, i.e. $\frac{dL_{eff}}{dE_\gamma}=L_{eA}
\frac{dN}{dE_\gamma}$. The number of events per unit time, $N_\tau$,
is given by the integral $N_\tau=\int \sigma(E_\gamma)  dL_{eff} $,
where $\sigma(E_\gamma)$ is the photonuclear cross section. Assuming
that the photonuclear cross section peaks at energy $E_0$ and using
the Thomas-Reiche-Kuhn (TRK) sum rule \cite{BHM02}, we can
approximate this integral by $N_\tau=\frac{dL_{eff}}{dE_0} \times
6\times 10^{-26}\frac{NZ}{A}$, where $dL_{eff}/dE$ is expressed in
units of cm$^{-2}$s$^{-1}$MeV$^{-1}$. The giant resonances exhaust
most part of the TRK sum rule and occur in nuclei at energies around
$E_0=15$ MeV. For 1 GeV electrons $dN(E_\gamma=E_0)/dE_\gamma\approx
6\times 10^{-3}$/MeV. With a luminosity of $L_{eA}=10^{25}$
cm$^{-2}$s$^{-1}$, one gets $\frac{dL_{eff}}{dE_0}\approx 6\times
10^{22}$ cm$^{-2}$MeV$^{-1}$s$^{-1}$ and a number of events
$N_\tau\approx4\times 10^{-3}NZ/A\approx10^{-3}A $ s$^{-1}$. Thus,
for medium mass nuclei, one expects thousands of events per day.
These estimates increase linearly with the accelerator luminosity,
$L_{eA}$, and show that studies of giant resonances in neutron-rich
nuclei is very promising at the proposed facilities
\cite{Haik05,Sud01}. Only a small fraction, of the order of
5\%-10\%, of the TRK sum-rule goes into the excitation of
soft-dipole modes \cite{AGB82}. However, these modes occur at a much
lower energy, $E_r\approx 1$ MeV, where the number of equivalent
photons (see figure 2) is at least one order of magnitude larger
than for giant resonance energies. Therefore, inelastic processes
leading to the excitation of soft dipole modes will be as abundant
as those for excitation of giant resonances. However, one has to
keep in mind that it is not clear if experiments with very
short-lived nuclei will be feasible at the proposed electron-ion
colliders.

\section{Dissociation of weakly-bound systems}

\subsection{One-neutron halo}

In this section I will consider the dissociation of a weakly-bound
(halo) nucleus from a bound state into a structureless continuum. I
calculate the matrix elements for the response function in eq.
\ref{photo} with a two-body model which has been used previously to
study Coulomb excitation of halo nuclei with relative success
\cite{BB86,BS92,Ots94,MOI95,KB96,TB04}.
The initial wavefunction can be written as $\Psi_{JM}=r^{-1}u_{ljJ}%
(r)\mathcal{Y}_{lJM}$, where $R_{ljJ}(r)$ is the radial wavefunction and
$\mathcal{Y}_{lJM}$ is a spin-angle function \cite{BD04}. The radial
wavefunction, $u_{ljJ}(r)$, can be obtained by solving the radial
Schr\"{o}dinger equation for a nuclear potential, $V_{Jlj}^{(N)}(r)$. Some
analytical insight may be obtained using a simple Yukawa form for an s-wave
initial wavefunction, $u_{0}(r)=A_{0}\exp(-\eta r)$, and a p-wave final
wavefunction, $\ u_{1}(r)=j_{1}(kr)\cos\delta_{1}-\ n_{1}(kr)\sin\delta_{1}$.
In these equations $\eta$\ is related to the neutron separation energy
$S_{n}=\hbar^{2}\eta^{2}/2\mu$, $\mu$\ is the reduced mass of the neutron +
core system, and $\hbar k=\sqrt{2\mu E_{r}}$, with $E_{r}$ being the final
energy of relative motion between the neutron and the core nucleus. $A_{0}$ is
the normalization constant of the initial wavefunction. The transition density
is given by $r^{2}\delta\rho_{if}\left(  r\right)  =e_{eff}A_{i}u_{i}%
(r)u_{f}(r)$, where $i$ and $f$ indices include angular momentum dependence
and $e_{eff}=-eZ_{c}/A$ is the effective charge of a neutron+core nucleus with
charge $Z_{c}$. The $E1$ transition integral $\mathcal{I}_{l_{i}l_{f}}%
=\int_{0}^{\infty}dr\ r^{3}\ \ \delta\rho_{if}\left(  r\right)  $ for the
wavefunctions described above yields%
\begin{align}
\mathcal{I}_{s\rightarrow p}  &  =e_{eff}\frac{2k^{2}}{\left(  \eta^{2}%
+k^{2}\right)  ^{2}}\left[  \cos\delta_{1}+\sin\delta_{1}\frac{\eta\left(
\eta^{2}+3k^{2}\right)  }{2k^{3}}\right] \nonumber\\
&  \simeq\frac{e_{eff}\hbar^{2}}{2\mu}\frac{2E_{r}}{\left(  S_{n}%
+E_{r}\right)  ^{2}}\left[  1+\left(  \frac{\mu}{2\hbar^{2}}\right)
^{3/2}\frac{\sqrt{S_{n}}\left(  S_{n}+3E_{r}\right)  }{-1/a_{1}+\mu r_{1}%
E_{r}/\hbar^{2}}\right]  , \label{isp}%
\end{align}
where the effective range expansion of the phase shift, $k^{2l+1}\cot
\delta\simeq-1/a_{l}+r_{l}k^{2}/2,$ was used in the second line of the above
equation. For $l=1$, $a_{1}$ is the \textquotedblleft scattering
volume\textquotedblright\ (units of length$^{3}$) and $r_{1}$ is the
\textquotedblleft effective momentum\textquotedblright\ (units of 1/length).
Their interpretation is not as simple as the $l=0$ effective range parameters.
Typical values are, e.g. $a_{1}=-13.82$ fm$^{-3}$ and $r_{1}=-0.419$ fm$^{-1}$
for n+$^{4}$He $p_{1/2}$-wave scattering and $a_{1}=-62.95$ fm$^{-3}$ and
$r_{1}=-0.882$ fm$^{-1}$ for n+$^{4}$He $p_{3/2}$-wave scattering \cite{Arn73}.

The energy dependence of eq. \ref{isp} has some unique features. As
shown in previous works \cite{BB86,BBH91,BS92}, the matrix elements
for electromagnetic response of weakly-bound nuclei present a small
peak at low energies, due to the proximity of the bound state to the
continuum. This peak is manifest in the response function of eq.
\ref{photo}:
\begin{equation}
\frac{dB(EL)}{dE}\propto\left\vert \mathcal{I}_{s\rightarrow p}\right\vert
^{2}\propto\frac{E_{r}^{L+1/2}}{\left(  S_{n}+E_{r}\right)  ^{2L+2}}.
\label{deel}%
\end{equation}
It appears centered at the energy \cite{BS92}
$E_{0}^{(EL)}\simeq\frac{L+1/2}{L+3/2}S_{n}$ for a generic electric
response of multipolarity $L$. For $E1$ excitations, the peak occurs
at $E_{0}\simeq3S_{n}/5$.

\begin{figure}[ptb]
\begin{center}
\includegraphics[
height=2.9464in,
width=3.1298in
]{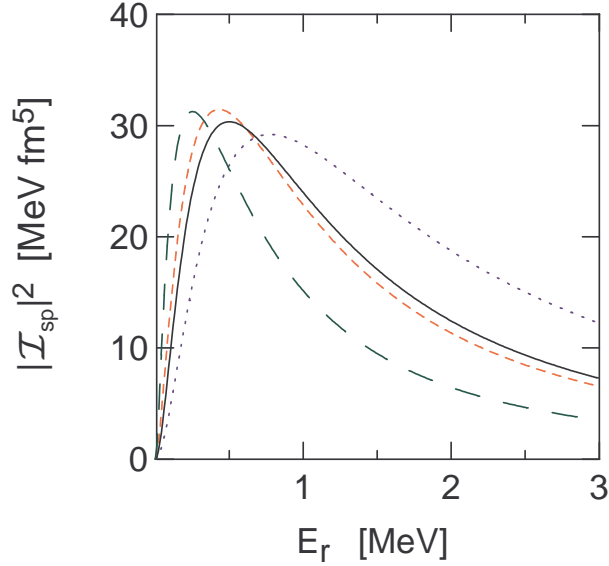}
\end{center}
\caption{(Color online) $\left\vert \mathcal{I}_{s\rightarrow
p}\right\vert ^{2}$ \ calculated using eq. \ref{isp}, assuming
$e_{eff}=e$, $A=11$ and $S_{n}=0.5$ MeV, as a function of $E_{r}$.
The long dashed curve corresponds to \ $a_{1}=-10$ fm$^{-3}$ and
$r_{1}=-0.5$ fm$^{-1}$, the dashed curve corresponds to \
$a_{1}=-50$ fm$^{-3}$ and $r_{1}=1$ fm$^{-1},$ the solid curve
corresponds to\ $a_{1}=$ $r_{1}=0$, and finally, the dotted curve
corresponds to\ $a_{1}=-10$ fm$^{-3}$ and $r_{1}=0.5$ fm$^{-1}$.}%
\label{fig3}%
\end{figure}

The second term inside brackets in eq. \ref{isp} is a modification
due to final state interactions. This modification may become
important, as shown in figure \ref{fig3}, where $\left\vert
\mathcal{I}_{s\rightarrow p}\right\vert ^{2}$ calculated with eq.
\ref{isp} is plotted as a function of $E_{r}$. Here, for simplicity,
I have assumed the values $e_{eff}=e$, $A=11$ and $S_{n}=0.5$ MeV.
This does not correspond to any known nucleus and is used to assess
the effect of the scattering length and effective range in the
transition matrix element. The long dashed curve corresponds to \
$a_{1}=-10$ fm$^{-3}$ and $r_{1}=-0.5$ fm$^{-1}$, the dashed curve
corresponds to \ $a_{1}=-50$ fm$^{-3}$ and $r_{1}=1$ fm$^{-1},$ the
solid curve corresponds to\ $a_{1}=$ $r_{1}=0$, and finally, the
dotted curve corresponds to\ $a_{1}=-10$ fm$^{-3}$ and $r_{1}=0.5$
fm$^{-1}.$ Although the effective range expansion is only valid for
small values of $E_{r}$, it is evident from the figure that the
matrix element is very sensitive to the effective range expansion
parameters.

The strong dependence of the response function on the effective
range expansion parameters makes it an ideal tool to study the
scattering properties of light nuclei which are of interest for
nuclear astrophysics. It is important to notice that the one-halo
has been studied in many experiments, e.g. for the case of $^{11}$Be
for which there are many data available (see refs.
\cite{Nak94,Pal03,Fuk04}). In these papers one can find a detailed
analysis of how the nuclear shell-model can explain the experimental
data, by fitting the spectroscopic factors for several
single-particle configurations. It is beyond the purpose of the
present paper to reproduce theses data, in view of the simple model
adopted above. The main goal of this section is to show the
relevance of final state interactions.

\subsection{Two-neutron halo}

\begin{figure}[ptb]
\begin{center}
\includegraphics[
height=2.2485in,
width=2.6117in
]{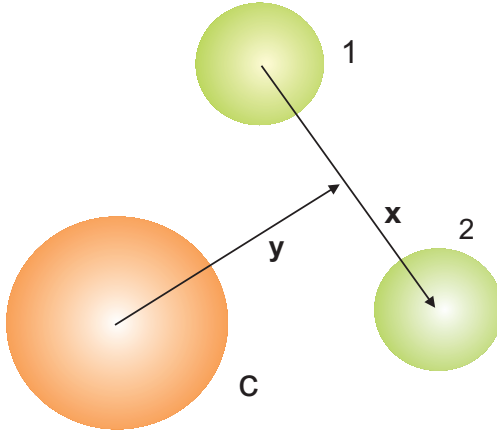}
\end{center}
\caption{(Color online) Jacobian coordinates (\textbf{x} and
\textbf{y}) for a three-body
system consisting of a core (c) and two nucleons (1 and 2). }%
\label{threeb}%
\end{figure}

Many weakly-bound nuclei, like $^{6}$He or $^{11}$Li, require a three--body
treatment in order to reproduce the electromagnetic response more accurately.
In a popular three-body model, the bound--state wavefunction in the center of
mass system is written as an expansion over hyperspherical harmonics (HH), see
e.g.~\cite{Zhu93},
\begin{equation}
\Psi\left(  \mathbf{x},\mathbf{y}\right)  =\frac{1}{\rho^{5/2}}\sum
_{KLSl_{x}l_{y}}\Phi_{KLS}^{l_{x}l_{y}}\left(  \rho\right)  \left[
\mathcal{J}_{KL}^{l_{x}l_{y}}\left(  \Omega_{5}\right)  \otimes\chi
_{S}\right]  _{JM}. \label{hhwf}%
\end{equation}
Here $\mathbf{x}$ and $\mathbf{y}$ are Jacobi vectors where (see
figure \ref{threeb}) $\mathbf{x}=\frac{1}{\sqrt{2}}\left(
\mathbf{r}_{1}-\mathbf{r}_{2}\right)$ and $y=\sqrt{\frac{2\left(
A-2\right)  }{A}}\left(
\frac{\mathbf{r}_{1}+\mathbf{r}_{2}}{2}-\mathbf{r}_{c}\right)$,
where A is the nuclear mass, $\mathbf{r}_{1}$ and $\mathbf{r}_{2}$
are the position of the nucleons, and $\mathbf{r}_{c}$ is the
position of the core. The hyperradius \ $\rho$ determines the size
of a three-body state: $\rho ^{2}=x^{2}+y^{2}$. The five angles
$\left\{ \Omega_{5}\right\}  $ include usual angles
$(\theta_{x},\phi_{x})$, $(\theta_{y},\phi_{y})$ which parametrize
the direction of the unit vectors $\widehat{\mathbf{x}}$ and
$\widehat{\mathbf{y}}$ and the hyperangle $\theta$, related by
$x=\rho \sin\theta$ and $y=\rho\cos\theta$, where
$0\leq\theta\leq\pi/2$.

The insertion of the three-body wavefunction, eq. \ref{hhwf}, into
the Schr\"odinger equation yields a set of coupled differential
equations for the hyperradial wavefunction
$\Phi_{KLS}^{l_{x}l_{y}}\left(  \rho\right) $. Assuming that the
nuclear potentials between the three particles are known, this
procedure yields the bound-state wavefunction for a three-body
system with angular momentum $J$.

In order to calculate the electric response we need the scattering
wavefunctions in the three-body model to calculate the integrals in
eq. \ref{photo}. One would have to use final wavefunctions with
given momenta, including their angular information. When the final
state interaction is disregarded these wavefunctions are three-body
plane waves \cite{Pus96,Chu93}. To carry out the calculations, the
plane waves can be expanded in products of hyperspherical harmonics
in coordinate and momentum spaces. However, since we are only
interested in the energy dependence of the response function, we do
not need directions of the momenta. Thus, instead of using plane
waves, I will use a set of final states which just include the
coordinate space and energy dependence.

I will also adopt an approach closely related to the work of Pushkin
et al. \cite{Pus96} (see also \cite{For02a,For02b}). For
weakly-bound systems having no bound subsystems the hyperradial
functions entering the expansion \ref{hhwf} behave asymptotically as
\cite{Mer74} $ \Phi_{a}\left(  \rho\right)
\longrightarrow\text{constant}\times\ \exp\left( -\eta\rho\right)$
as $\rho\longrightarrow \infty$,
where the two-nucleon separation energy is related to $\eta$ by $S_{2n}%
=\hbar^{2}\eta^{2}/\left(  2m_{N}\right)  $. This wavefunction has
similarities with the two-body case, when $\rho$ is interpreted as the
distance $r$ between the core and the two nucleons, treated as one single
particle. But notice that the mass $m_{N}$ would have to be replaced by
$2m_{N}$ if a simple two-body (the dineutron-model \cite{BB88,BBH91}) were
used for $^{11}$Li or $^{6}$He.

\begin{figure}[ptb]
\begin{center}
\includegraphics[
height=2.6187in,
width=3.3347in
]{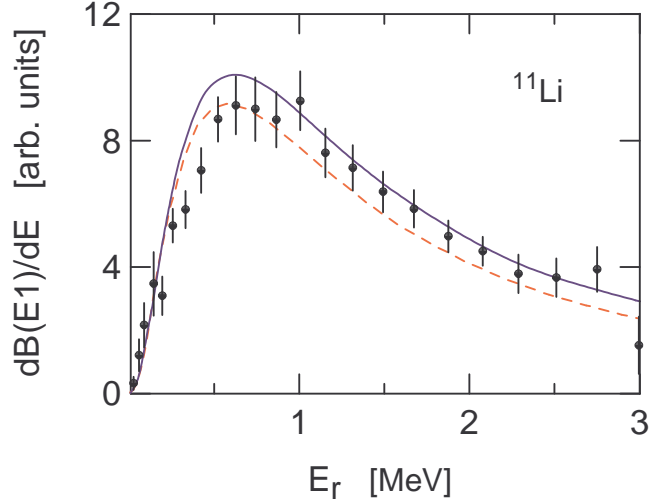}
\end{center}
\caption{(Color online) Comparison between the calculation of the
response function (in arbitrary units) with eqs. \ref{ie1} and
\ref{dbde3b}, using $\delta_{nn}=0$ and $\delta_{nc}=0$, (dashed
line), or including the effects of final state interactions
(continuous line). The experimental data are from ref.
\cite{Shi95}. }%
\label{fig4}%
\end{figure}

Since only the core carries charge, in a three-body model the $E1$
transition operator is given by $M\sim yY_{1M}\left(
\widehat{\mathbf{y}}\right)  $ for the final state (see also
\cite{Chu93}). The $E1$ transition matrix element is obtained by a
sandwich of this operator between $\Phi_{a}\left(  \rho\right)
/\rho^{5/2}$ and scattering wavefunctions. In ref. \cite{Pus96} the
scattering states were taken as plane waves. I will use distorted
scattering states,
leading to the expression%
\begin{equation}
\mathcal{I}\left(  E1\right)  =\int dxdy\frac{\Phi_{a}\left(  \rho\right)
}{\rho^{5/2}}\ y^{2}x u_{p}\left(  y\right)  u_{q}\left(  x\right)  ,
\label{ie1}%
\end{equation}
where $u_{p}\left(  y\right)  =j_{1}\left(  py\right)  \cos\delta_{nc}%
-n_{1}\left(  py\right)  \cos\delta_{nc}$ is the core-neutron
asymptotic continuum wavefunction, assumed to be a $p$-wave, and
$u_{q}\left(  x\right) =j_{0}\left(  qx\right)
\cos\delta_{nn}-n_{0}\left(  qx\right)  \cos \delta_{nn}$ is the
neutron-neutron asymptotic continuum wavefunction, assumed to be an
$s$-wave. The relative momenta are given by
$\mathbf{q}=\frac{1}{\sqrt{2}}\left(
\mathbf{q}_{1}-\mathbf{q}_{2}\right)$, and
$\mathbf{p}=\sqrt{\frac{2\left( A-2\right)  }{A} }\left(
\frac{\mathbf{k}_{1}+\mathbf{k}_{2}}{2}-\mathbf{k}_{c}\right)$.

The $E1$ strength function is proportional to the square of the matrix element
in eq. \ref{ie1} integrated over all momentum variables, except for the total
continuum energy $E_{r}=\hbar^{2}\left(  q^{2}+p^{2}\right)  /2m_{N}$. This
procedure gives%
\begin{equation}
\frac{dB\left(  E1\right)  }{dE_{r}}=\text{constant }\times\int\left\vert
\mathcal{I}\left(  E1\right)  \right\vert ^{2}E_{r}^{2}\cos^{2}\Theta\sin
^{2}\Theta d\Theta d\Omega_{q}d\Omega_{p}, \label{dbde3b}%
\end{equation}
where $\Theta=\tan^{-1}\left(  q/p\right)  $.

The $^{1}$S$_{0}$ phase shift in neutron-neutron scattering is
remarkably well reproduced up to center of mass energy of order of 5
MeV by the first two
terms in the effective-range expansion $k\cot\delta_{nn}\simeq-1/a_{nn}%
+r_{nn}k^{2}/2.$ Experimentally these parameters are determined to be
$a_{nn}=-23.7$ fm and $r_{nn}=2.7$ fm$.$ The extremely large (negative) value
of the scattering length implies that there is a virtual bound state in this
channel very near zero energy. The p-wave scattering in the n-$^{9}$Li
($^{10}$Li) system appears to have resonances at low energies \cite{Thoe99}.
I assume that this phase-shift can be described by the resonance relation%
$\sin\delta_{nc}=({\Gamma/2})/{\sqrt{\left(  E_{r}-E_{R}\right)  ^{2}%
+\Gamma^{2}/4}}$, with $E_{R}=0.53$ MeV and $\Gamma=0.5$ MeV
\cite{Thoe99}.

Most integrals in eqs. \ref{ie1} and \ref{dbde3b} can be done analytically,
leaving two remaining integrals which can only be performed numerically. The
result of the calculation is shown in figure \ref{fig4}. The dashed line was
obtained using $\delta_{nn}=0$ and $\delta_{nc}=0$, that is, by neglecting
final state interactions. The continuous curve includes the effects of final
state interactions, with $\delta_{nn}$ and $\delta_{nc}$ parametrized as
described above. The experimental data are from ref. \cite{Shi95}. The data
and theoretical curves are given in arbitrary units. Although the experimental
data is not perfectly described by either one of the results, it is clear that
final state interactions are of extreme relevance.

As pointed out in ref. \cite{Pus96}, the $E1$ three-body response
function of $^{11}$Li can still be described by an expression
similar to eq. \ref{deel}, but with different powers. Explicitly,
${dB\left(  E1\right)  }/{dE_{r}}\propto{E_{r}^{3}}/{\left(
S_{2n}^{eff}+E_{r}\right)  ^{11/2}}$. Instead of $S_{2n}$, one has
to use an effective $S_{2n}^{eff}=aS_{2n}$, with $a\simeq1.5$. With
this approximation, the peak of the strength function in the
three-body case is situated at about three times higher energy than
for the two-body case, eq. \ref{deel}. In the
three-body model, the maximum is thus predicted at $E_{0}^{(E1)}%
\simeq1.8S_{2n}$, which fits the experimentally determined peak position for
the $^{11}$Li $E1$ strength function very well \cite{Pus96}. It is thus
apparent that the effect of three-body configurations is to widen and to shift
the strength function $dB\left(  E1\right)  /dE$ to higher energies.

It is worthwhile mentioning that the data presented in figure
\ref{fig4} and of other experiments \cite{Iek93,Zin97} is different
in form and magnitude of the more recent experiment of Nakamura et
al. \cite{Nak06}. The reason for the discrepancy is attributed to an
enhanced sensitivity in the experiment of ref. \cite{Nak06} to low
relative energies below $E_{rel}=0.5$ MeV compared to previous
experiments. Also, this recent experiment agrees very well with the
nn-correlated model of Esbensen and Bertsch \cite{EB92}. This
theoretical model is different than the model presented in this
section in many aspects. In principle, the three-body models should
be superior, as they include the interactions between the
three-particles without any approximation. For example,  ref.
\cite{EB92} use a simplified interaction between the two-neutrons.
On the other hand, they include the many-body effects, e.g. the
Pauli blocking of the occupied states in the core. It is not well
known the reason why the data of ref. \cite{Nak06} is better
described with the model of ref. \cite{EB92} than traditional 3-body
models.

\section{Collective excitations: the pigmy resonance}

\subsection{The hydrodynamical model}

We have seen that the energy position where the soft dipole response
peaks depends upon the few body model adopted. Except for a two-body
resonance in $^{10}$Li, there was no reference to a resonance in the
continuum. The peak in the response function can be simply explained
by the fact that it has to grow from zero at low energies and return
to zero at large energies. In few-body, or cluster, models, the form
of the bound-state wavefunctions and the phase space in the
continuum determine the position of the peak in the response
function. Few-body resonances will lead to more peaks.

Now I shall consider the case in which a collective resonance is
present. As with giant dipole resonances (GDR) in stable nuclei, one
believes that pygmy resonances at energies close to the threshold
are present in halo, or neutron-rich, nuclei. This was proposed by
Suzuki et al. \cite{SIS90} using the hydrodynamical model for
collective vibrations. The possibility to explain the soft dipole
modes (figure \ref{fig4}) in terms of direct breakup, has made it
very difficult to clearly identify the signature of pygmy resonances
in light exotic nuclei.

The hydrodynamical model, first suggested by Goldhaber and Teller
\cite{GT48} and by Steinwedel and Jensen \cite{SJ50} needs
adjustments to explain collective response in light, neutron-rich,
nuclei. Because clusterization in light nuclei exists, not all
neutrons and protons can be treated equally. The necessary
modifications are straight-forward and discussed next. To my
knowledge, the radial dependence of the transition densities in the
hydrodynamical model for light, neutron-rich, nuclei has not been
discussed in the literature. I will use the method of Myers et al.
\cite{Mye77}, who considered collective vibrations in nuclei as an
admixture of Goldhaber-Teller and Steinwedel-Jensen modes.

\begin{figure}[ptb]
\begin{center}
\includegraphics[
height=2.3082in, width=2.8911in ]{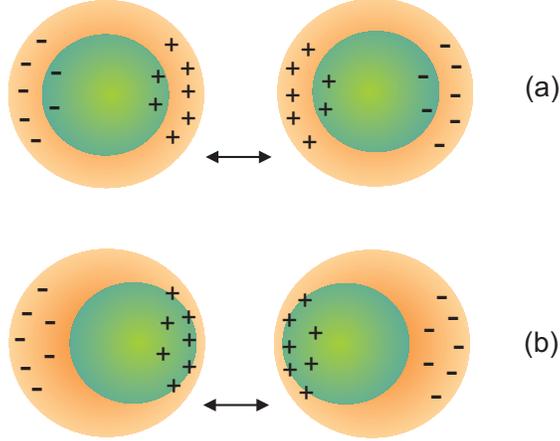}
\end{center}
\caption{(Color online) Hydrodynamical model for collective nuclear
vibrations in halo nuclei. The (a) Steinwedel-Jensen (SJ) mode and
the (b) Goldhaber-Teller (GT)
mode are shown separately.}%
\label{gtsj}%
\end{figure}

When a collective vibration of protons against neutrons is present
in a nucleus with charge (neutron) number $Z$ ($N$), the neutron and
proton fluids are displaced with respect to each other by
$d_{1}=\alpha_{1}R$ and each of the fluids are displaced from the
origin (center of mass of the system) by $d_{p}=Nd_{1}/A$ and $d_{n}%
=-Zd_{1}/A$. This leaves the center of mass fixed and one gets for
the dipole moment $D_{1}=Zed_{p}=\alpha _{1}NZeR/A$. The GT model
assumes that the restoring force is due to the increase of the
nuclear surface which leads to an extra energy proportional to
$A^{2/3}$. In this model, the inertia is proportional to $A$ and the
excitation energy is consequently given by $E_{x}\propto\sqrt{A^{2/3}%
/A}=A^{-1/6}$.

For light, weakly-bound nuclei, it is more appropriate to assume
that the neutrons inside the core ($A_{c},Z_{c}$) vibrate in phase
with the protons. The neutrons and protons in the core are tightly
bound. An overall displacement among them requires energies of the
order of 10-20 MeV, well above that of the soft dipole modes. The
dipole moment becomes $\mathbf{D}_{1} =e\mathbf{d}_{1}{\left(  Z_{c}%
A-ZA_{c}\right)  }/{A}  =Z_{eff}^{(1)}e\mathbf{d}_{1}$, where
$\mathbf{d}_{1}$ is a vector connecting the center of mass of the
two fluids (core and excess neutrons).  We see that the dipole
moment is now smaller than before because the
effective charge changes from $NZ/A$ in the case of the GDR to $Z_{eff}%
^{(1)}=$ $\left(  Z_{c}A-ZA_{c}\right)  /A$. This effective charge is zero if
$Z_{c}A=ZA_{c}$ and no pigmy resonance is possible in this model, only the
usual GDR.

Figure \ref{gtsj} shows a schematic representation of the hydrodynamical model
for collective nuclear vibrations in a halo nucleus, as considered here. Part
(a) of the figure shows the Steinwedel-Jensen (SJ) mode in which the total
matter density of both the core and the halo nucleons do not change locally.
Only the local ratio of the neutrons and protons changes. Part (b) of the
figure shows a particular case of the Goldhaber-Teller (GT) mode, in which the
core as a whole moves with respect to the halo nucleons.

\begin{figure}[ptb]
\begin{center}
\includegraphics[
natheight=3.589000in,
natwidth=4.678600in,
height=3.0597in,
width=3.9825in
]{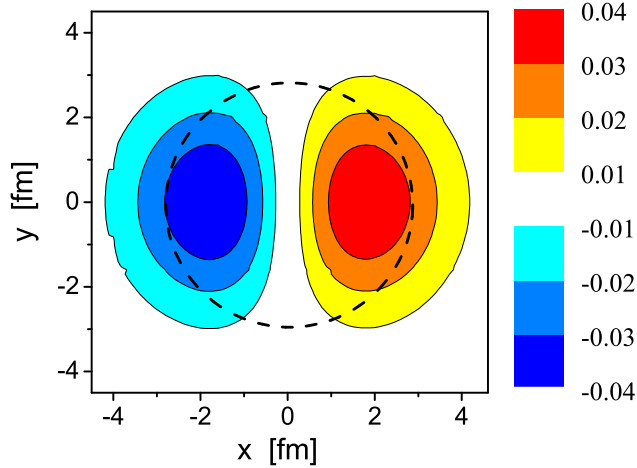}
\end{center}
\caption{(Color online) Contour plot of the nuclear transition
density in the hydrodynamical model consisting of a mixture of GT
and SJ vibrations. The darker areas represent the larger values of
the transition density in a nucleus which has an average radius
represented by the dashed circle. The legend on the right
displays the values of the transition density within each contour limit. }%
\label{pygmy}%
\end{figure}

For spherically symmetric densities, the transition density in the GT mode can
be calculated from $\delta\rho_{p}=\rho_{p}\left(  \left\vert \mathbf{r}%
-\mathbf{d}_{p}\right\vert \right)  -\rho_p\left(  \mathbf{r}\right)
$, where $\rho_p$ is the charge density. Using $d_{1}\ll R$, it is
straight-forward to show that  $ \delta\rho_{p}^{(1)}\left(
\mathbf{r}\right) =\delta\rho_{p}^{(1)}\left(  r\right) Y_{10}\left(
\widehat{\mathbf{r}}\right)$ , where
\begin{equation}
\delta\rho_{p}^{(1)}\left(  r\right)
=\sqrt{\frac{4\pi}{3}}Z_{eff}\alpha _{1}R\frac{d\rho_{0}}{dr},
\end{equation}
and $\rho_0$ is the ground state matter density.

In the Steinwedel-Jensen (SJ) mode, the local variation of the
density of protons is found to be $\delta\rho_{p}^{(2)}\left(
\mathbf{r}\right)  =\delta \rho_{p}^{(2)}\left(  r\right)
Y_{10}\left( \widehat{\mathbf{r}}\right)$,
where%
\begin{equation}
\delta\rho_{p}^{(2)}\left(  r\right)  =\sqrt{\frac{4\pi}{3}}Z_{eff}%
^{(2)}\alpha_{2}Kj_{1}\left(  kr\right)  \rho_{0}\left(  r\right)  ,
\end{equation}
where $K=9.93$. If the proton and neutron content of the core does
not change \cite{SIS90},
the effective charge number in the SJ mode is given by $Z_{eff}^{(2)}%
=Z^{2}(N-N_{c})/A(Z+N_{c})$.

The transition density at a point $\mathbf{r}$ from the
center-of-mass of the nucleus is a combination of the SJ and GT
distributions and is given by $ \delta\rho_{p}\left(
\mathbf{r}\right)     =\delta\rho_{p}\left(  r\right)  Y_{10}\left(
\widehat{\mathbf{r}}\right)$,
where%
\begin{equation}
\delta\rho_{p}\left(  r\right)  =\sqrt{\frac{4\pi}{3}}R\left\{  Z_{eff}%
^{(1)}\alpha_{1}\frac{d}{dr}+Z_{eff}^{(2)}\alpha_{2}\frac{K}{R}j_{1}\left(
kr\right)  \right\}  \rho_{0}(r). \label{transdsjgt}%
\end{equation}
Changes can be accommodated in these expressions to account for the different
radii of the proton and neutron densities.

Figure \ref{pygmy} shows the contour plot, in arbitrary units, of
the nuclear transition density in the hydrodynamical model,
consisting of a mixture of GT and SJ vibrations. The darker areas
represent the larger values of the transition density in a nucleus
which has an average radius represented by the dashed circle. In
this particular case, I have used the HF density \cite{Ber06,SB97}
for $^{11}$Li, and a radius $R=3.1$ fm. The parameters $\alpha_{1}$
and $\alpha_{2}$ were chosen so that $Z_{eff}^{(1)}\alpha
_{1}=Z_{eff}^{(2)}\alpha_{2}$, i.e$.$ a symmetric mixture of the SJ
and GT modes.

\begin{figure}[ptb]
\begin{center}
\includegraphics[
height=2.6333in,
width=3.4688in
]{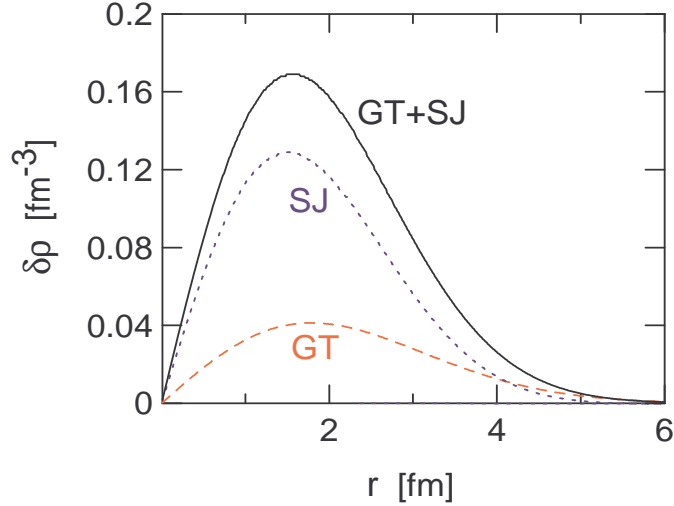}
\end{center}
\caption{(Color online) Hydrodynamical transition densities for
$^{11}$Li and three different assumptions for the SJ+GT admixtures,
according to eq. \ref{transdsjgt}. The dashed curve is for a GT
oscillation mode, with the core vibrating against the halo neutrons,
with effective charge number $Z_{eff}^{(1)}=6/11$, radius $R=3.1$
fm, and $\alpha_{1}=1$. The dotted curve is for an SJ oscillation
mode, with effective charge number $Z_{eff}^{(2)}=2/11$, and
$\alpha_{2}=1$.
The solid curve is their sum.}%
\label{trand}%
\end{figure}

Figure \ref{trand} shows the transition densities for $^{11}$Li for
three different assumptions of the SJ+GT admixtures, according to
eq. \ref{transdsjgt}. The dashed curve is for a GT oscillation mode,
with the core vibrating against the halo neutrons, with effective
charge number $Z_{eff}^{(1)}=6/11$, radius $R=3.1$ fm, and
$\alpha_{1}=1$. The dotted curve
is for an SJ oscillation mode, with effective charge number $Z_{eff}%
^{(2)}=2/11$, and $\alpha_{2}=1$. The solid curve is their sum.
Notice that the transition densities are peaked at the surface, but
at a
radius smaller than the adopted \textquotedblleft rms\textquotedblright%
\ radius $R=3.1$ fm.

The liquid drop model predicts an equal admixture of SJ+GT
oscillation modes for large nuclei  \cite{Mye77}. The contribution
of the SJ oscillation mode decreases with decreasing mass number,
i.e. $\alpha_2 \longrightarrow 0  $ as $A\longrightarrow0$. This is
even more probable in the case of halo nuclei, where a special type
of GT mode (oscillations of the core against the halo nucleons) is
likely to be dominant. For this special collective motion an
approach different than those used in refs. \cite{Mye77} and
\cite{SIS90} has to be considered.  The resonance energy formula
derived by Goldhaber and Teller \ \cite{GT48} changes to%
\begin{equation}
E_{PR}=\left(  \frac{3\varphi\hbar^{2}}{2aRm_{N}A_{r}}\right)  ^{1/2},
\label{EPGDR}%
\end{equation}
where $A_{r}=A_{c}\left(  A-A_{c}\right)  /A$ and $a$ is the length within
which the interaction between a neutron and a nucleus changes from a
zero-value outside the nucleus to a high value inside, i.e. $a$ is the size of
the nuclear surface. $\varphi$ is the energy needed to extract one neutron
from the proton environment.

Goldhaber and Teller \cite{GT48} argued that in a heavy stable
nucleus $\varphi$ is not the binding energy of the nucleus, but the
part of the potential energy due to the neutron-proton interaction.
It is proportional to the asymmetry energy. In the case of
weakly-bound nuclei this picture changes and it is more reasonable
to associate $\varphi$ to the separation energy of the valence
neutrons, $S$. I will use $\varphi=\beta S$, with a parameter
$\beta$ which is expected to be of order of one. Since for halo
nuclei the product $aR$ \ is proportional to $S^{-1},$ we obtain the
proportionality $E_{PR}\propto S$. Using eq. \ref{EPGDR} for
$^{11}$Li , with \ $a=1$ fm, $R=3$ fm and $\varphi=S_{2n}=0.3$ MeV,
we get $E_{PR}=1.3$ MeV. Considering that the pygmy resonance will
most probably decay by particle emission, one gets $E_{r}\simeq1$
MeV for the kinetic energy of the fragments, which is within the
right ballpark (see figure \ref{fig4}).

Both the direct dissociation model and the hydrodynamical model
yield a bump in the response function proportional to $S$, the
valence nucleon(s) separation energy. In the direct dissociation
model the width of the response function obviously depends on the
separation energy. But it also depends on the nature of the model,
i.e. if it is a two-body model, like the model often adopted for
$^{11}$Be or $^{8}$B, or a three-body model, appropriate for
$^{11}$Li and $^{6}$He. In the two-body model the phase-space
depends on energy as $\rho\left( E\right)  \propto
d^{3}p/dE\propto\sqrt{E}$, while in the three-body model $\rho\left(
E\right)  \sim E^{2}$. This explains why the peak of figure
\ref{fig4} is pushed toward higher energy values, as compared to the
prediction of eq. \ref{deel}. It also explains the larger width of
$dB/dE$ obtained in three-body models. In the case of the pigmy
resonance model, this question is completely open.

The hydrodynamical model predicts \cite{Mye77} for the width of the
collective mode $\Gamma=\hbar\overline{\mathrm{v}}/R$, where
$\overline{\mathrm{v}}$ \ is the average velocity of the nucleons
inside the nucleus. This relation can be derived by assuming that
the collective vibration is damped by the incoherent collisions of
the nucleons with the walls of the nuclear potential well during the
vibration cycles (piston model). Using
$\overline{\mathrm{v}}=3\mathrm{v}_{F}/4$, where
$\mathrm{v}_{F}=\sqrt{2E_{F}/m_{N}}$ is the Fermi velocity, with
$E_{F}=35$ MeV and $R=6$ fm, one gets $\Gamma\simeq6$ MeV. This is
the typical energy width a giant dipole resonance state in a heavy
nucleus. In the case of neutron-rich light nuclei
$\overline{\mathrm{v}}$ is not well defined. There are two average
velocities: one for the nucleons in the core, $\overline
{\mathrm{v}}_{c}$, and another for the nucleons in the skin, or
halo, of the nucleus, $\overline{\mathrm{v}}_{h}$. One is thus
tempted to use a
substitution in the form $\overline{\mathrm{v}}=\sqrt{\overline{\mathrm{v}%
}_{c}\overline{\mathrm{v}}_{h}}.$ Following ref. \cite{BMc93}, the width of
momentum distributions of core fragments in knockout reactions, $\sigma_{c}$,
is related to the Fermi velocity of halo nucleons by $\mathrm{v}_{F}%
=\sqrt{5\sigma_{c}^{2}}/m_{N}$. Using this expression with $\sigma_{c}%
\simeq20$ MeV/c, we get $\Gamma=5$ \ MeV (with $R=3$ fm). This value
is also not in discordance with experiments (see figure \ref{fig4}).

Better microscopic models, e.g. those based on random phase
approximation (RPA) calculations \cite{BF90,Te91} are necessary to
study pigmy resonances. The halo nucleons have to be treated in an
special way to get the response at the right energy position, and
with approximately the right width \cite{Te91,SB97}. Electron
scattering will provide a unique opportunity to clarify this issue
due to its better resolution over Coulomb excitation.

\subsection{Total inelastic cross sections}

One might argue that the total breakup cross section would be a good
signature for discerning direct dissociation versus the dissociation
through the excitation of a pigmy collective vibration. The trouble
is that the energy weighted sum rule for both cases are
approximately of the same magnitude \cite{BBH91,SIS90}. This can be
shown by using the electric dipole strength function in the direct
breakup
model of ref. \cite{BBH91}, namely%
\begin{equation}
\frac{dB(E1)}{dE}=\mathcal{C}\frac{3\hbar e^{2}Z_{eff}^{2}}{\pi^{2}\mu}%
\frac{\sqrt{S_{n}}\left(  E-S_{n}\right)  ^{3/2}}{E^{4}},\label{dbe1de}%
\end{equation}
where $E=E_{r}+S_{n}$ is the total excitation energy. $\mathcal{C}$
is a constant of the order of one, accounting for the corrections to
the wavefunction used in ref. \cite{BBH91}.

The sum rule for dipole excitations, $S_{1}\left(  E1\right)
=\int_{S_n}^\infty dE\ E\ \frac{dB\left( E1\right)}{dE}$, is
\begin{equation}
S_{1}\left(  E1\right)  =\mathcal{C}\left(  \frac{9}{8\pi}\right)  \frac
{\hbar^{2}e^{2}}{\mu}Z_{eff}^{2},\label{srule}%
\end{equation}
with $Z_{eff}^2=\left(  Z_{c}A-ZA_{c}\right)  ^{2}/[AA_{c}\left(  A-A_{c}%
\right)  ]$. This is the same (with $\mathcal{C}=1$) as eq. 1 of
ref. \cite{AGB82}, which is often quoted as the standard value to
which models for the nuclear response in the region of pigmy
resonance should be compared to. The response function in eq.
\ref{dbe1de}, with $\mathcal{C}=1$, therefore exhausts 100\% of the
so-called cluster sum rule  \cite{AGB82}. The total cross section
for electron breakup of weakly-bound systems is roughly proportional
to $S_{1}$. This assertion can be easily verified by using eqs.
\ref{EPA} and \ref{dbe1de}, assuming that the logarithmic dependence
of the virtual photon numbers on the energy $E\equiv E_{\gamma}$ can
be factored out of the integral in eq. \ref{EPA}.

The dipole strength of the pigmy dipole resonance is given by the
same equation \ref{srule}. The constant $\mathcal{C}$ is still of
the order of unity, but not necessarily the same as in eq.
\ref{dbe1de} and the effective charges are also different. For the
Goldhaber-Teller pigmy dipole model  the effective charge is given
by $Z_{eff}^{(1)}=$ $\left(  Z_{c}A-ZA_{c}\right)  /A$, whereas for
the Steinwedel-Jensen it is
$Z_{eff}^{(2)}=(Z^{2}/A)(N-N_{c})/(Z+N_{c})$. Assuming that the
Goldhaber-Teller mode prevails, one gets the simple prediction for
the ratio between the cross sections for
direct breakup versus excitation of a pigmy collective mode:%
\begin{equation}
\frac{\sigma^{direct}}{\sigma^{pigmy}}=\mathcal{C}\frac{A
}{A_{c}\left(  A-A_{c}\right)  }.
\end{equation}
For $^{11}$Li this ratio is 11$\mathcal{C}$/18 while for $^{11}$Be
it is 11$\mathcal{C}$/10.

\section{Summary and Conclusions}

I have studied the feasibility to determine low energy excitation
properties of light, exotic, nuclei from experimental data on
inelastic electron scattering. It was shown that for the conditions
attained in the electron-ion collider mode, the electron scattering
cross sections are directly proportional to photonuclear processes
with real photons. This proportionality is lost when larger
scattering angles, and larger ratio of the excitation energy to the
electron energy, $E_{\gamma}/E$, are involved.

One of the important issues to be studied in future electron-ion
colliders is the nuclear response at low energies. This response can
be modeled in two ways: by a (a) direct breakup and by a (a)
collective excitation. We have shown that in the case of direct
breakup the response function will depend quite strongly on the
final state interaction. This may become a very useful technique to
obtain phase-shifts, or effective-range expansion parameters, of
fragments far from the stability line. In the case of collective
excitations, a variant of the Goldhaber-Teller and Steinwedel-Jensen
model was used to obtain the transition densities in halo nuclei.

The pygmy resonance lies above the neutron emission threshold, effectively
precluding its observation in ($\gamma$,$\gamma^{\prime}$) experiments on very
neutron-rich nuclei. Nonetheless, electron scattering experiments will probe
the response function under several conditions, including different bombarding
energies, different scattering angles, etc. The study of pygmy resonances and
of final state interactions will certainly be an important line of
investigation in these facilities.

\acknowledgments{The author is grateful to Haik Simon and Toshimi
Suda for useful discussions. This work was supported by the
U.\thinspace S.\ Department of Energy under grant No.
DE-FG02-04ER41338.}

\end{document}